\documentclass[iop]{emulateapj}

\usepackage{hyperref}
\usepackage{xcolor}
\usepackage{url}
\usepackage{amssymb}
\hypersetup{
    colorlinks,
    linkcolor={red!50!black},
    citecolor={blue!50!black},
    urlcolor={black} 
}

\newcommand{\feh}{\ensuremath{[\mbox{Fe}/\mbox{H}]}}

\shorttitle{Thermal Spectrum of HD 88133 b}
\shortauthors{Piskorz et al.}

\begin{document}

\title{Evidence for the Direct Detection of the Thermal Spectrum of \\ the Non-Transiting Hot Gas Giant HD 88133 b}

\author{Danielle Piskorz\altaffilmark{1}, Bj{\"o}rn Benneke\altaffilmark{1}, Nathan R. Crockett\altaffilmark{1}, Alexandra C. Lockwood\altaffilmark{1,2}, \\Geoffrey A. Blake\altaffilmark{1}, Travis S. Barman\altaffilmark{3}, Chad F. Bender\altaffilmark{4,5}, Marta L. Bryan\altaffilmark{6}, John S. Carr\altaffilmark{7},\\Debra A. Fischer\altaffilmark{8}, Andrew W. Howard\altaffilmark{6}, Howard Isaacson\altaffilmark{10}, John A. Johnson\altaffilmark{11}}

\altaffiltext{1}{Division of Geological and Planetary Sciences, California Institute of Technology, Pasadena, CA 91125}
\altaffiltext{2}{King Abdullah University of Science and Technology, Thuwal Saudi Arabia}
\altaffiltext{3}{Lunar and Planetary Laboratory, University of Arizona, Tuscon, AZ 85721}
\altaffiltext{4}{Department of Astronomy and Astrophysics, Pennsylvania State University, University Park, PA 16802}
\altaffiltext{5}{Center for Exoplanets and Habitable Worlds, Pennsylvania State University, University Park, PA 16802}
\altaffiltext{6}{Cahill Center for Astronomy and Astrophysics, California Institute of Technology, Pasadena, CA 91125}
\altaffiltext{7}{Naval Research Laboratory, Washington, DC 20375}
\altaffiltext{8}{Department of Astronomy, Yale University, 260 Whitney Avenue, New Haven, CT 0651}
\altaffiltext{9}{Institute for Astronomy, University of Hawaii at Manoa, Honolulu, HI 96822}
\altaffiltext{10}{Department of Astronomy, University of California Berkeley, Berkeley, CA 94720}
\altaffiltext{11}{Harvard-Smithsonian Center for Astrophysics; Institute for Theory and Computation, Cambridge, MA 02138}

\begin{abstract}
We target the thermal emission spectrum of the non-transiting gas giant HD 88133 b with high-resolution near-infrared spectroscopy, by treating the planet and its host star as a spectroscopic binary. For sufficiently deep summed flux observations of the star and planet across multiple epochs, it is possible to resolve the signal of the hot gas giant's atmosphere compared to the brighter stellar spectrum, at a level consistent with the aggregate shot noise of the full data set. To do this, we first perform a principal component analysis to remove the contribution of the Earth's atmosphere to the observed spectra. Then, we use a cross-correlation analysis to tease out the spectra of the host star and HD 88133 b to determine its orbit and identify key sources of atmospheric opacity. In total, six epochs of Keck NIRSPEC \textit{L} band observations and three epochs of Keck NIRSPEC \textit{K} band observations of the HD 88133 system were obtained. Based on an analysis of the maximum likelihood curves calculated from the multi-epoch cross correlation of the full data set with two atmospheric models, we report the direct detection of the emission spectrum of the non-transiting exoplanet HD 88133 b and measure a radial projection of the Keplerian orbital velocity of 40 $\pm$ 15 km/s, a true mass of 1.02$^{+0.61}_{-0.28}M_J$, a nearly face-on orbital inclination of 15${^{+6}_{-5}}^{\circ}$, and an atmosphere opacity structure at high dispersion dominated by water vapor. This, combined with eleven years of radial velocity measurements of the system, provides the most up-to-date ephemeris for HD 88133.
\end{abstract}


\section{Introduction}
Since the discovery of 51 Peg b \citep{Mayor95}, the radial velocity (RV) technique has proven indispensable for exoplanet discovery. Hundreds of exoplanets have been revealed by measuring the Doppler wobble of the exoplanet host star \citep{Wright2012}, principally at visible wavelengths. To first order, the RV method yields the period and the minimum mass ($M\sin(i)$) of the orbiting planet. In order to complete the characterization of a given exoplanet, one would want to measure its radius and constrain its atmospheric constituents. Traditionally, this information is accessible only if the planet transits its host star with respect to our line of sight via transmission or secondary eclipse photometry. Successes with these techniques have resulted in the detections of water, carbon monoxide, and methane on the hottest transiting gas giants \citep{Madhu2014}. These gas giants orbit their host stars in days, are known as hot Jupiters, and have an occurence rate of only 1$\%$ in the exoplanet population \citep{Wright2012}. Broadband spectroscopic measurements of transiting hot Jupiter atmospheres are rarely able to resolve molecular bands, let alone individual lines, creating degeneracies in the solutions for atmospheric molecular abundances.

High-resolution infrared spectroscopy has recently provided another route to the study of exoplanet atmospheres, one applicable to transiting and non-transiting planets alike. Such studies capitalize on the Doppler shift between the stellar and planet lines, allowing them to determine the atmosphere compositions, true masses, and inclinations of various systems. With the CRyogenic InfraRed Echelle Spectrograph (CRIRES) at the VLT, \cite{Snellen2010} provided a proof-of-concept of this technique and detected carbon monoxide in the atmosphere of the transiting exoplanet HD 209458 b consistent with previous detections using Hubble Space Telescope data \citep{Swain2009}. By detecting the radial velocity variation of a planet's atmospheric lines in about six hours of observations on single nights, further work has detected the dayside and nightside thermal spectra of various transiting and non-transiting hot Jupiters, reporting detections of water and carbon monoxide, as well as the presence of winds and measurements of the length of day \citep{Snellen2010, Rodler2012, Snellen2014,Brogi2012, Brogi2013, Brogi2014, Brogi2016, Birkby2013,  deKok2013, Schwarz2015}. With HARPS, \cite{Martins2015} recently observed the reflected light spectrum of 51 Peg b in a similar manner, combining 12.5 hours of data taken over seven nights when the full dayside of the planet was observable.

\cite{Lockwood2014} studied the hot Jupiter tau Boo b using Keck NIRSPEC (Near InfraRed SPECtrometer), confirmed the CRIRES measurement of the planet's Keplerian orbital velocity, and detected water vapor in the atmosphere of a non-transiting exoplanet for the first time. NIRSPEC was used to observe an exoplanet's emission spectrum over $\sim$2-3 hours each night across multiple epochs, in order to capture snapshots of the planet's line-of-sight motion at distinct orbital phases. In combination with the many orders of data provided by NIRSPEC's cross dispersed echelle format and the multitude of hot Jupiter emission lines in the infrared, \cite{Lockwood2014} achieved sufficient signal-to-noise to reveal the orbital properties of tau Boo b via the Doppler shifting of water vapor lines in its atmosphere.

Here, we continue our Keck NIRSPEC direct detection program with a study of the emission spectrum of the hot gas giant HD 88133 b, a system that allows us to test the brightness limits of this method and develop a more robust orbital dynamics model that can be applied to eccentric systems. In Section~\ref{RVs} we present new (stellar) radial velocity (RV) observations of HD 88133 and an updated ephemeris. In Section~\ref{methods} we outline our NIRSPEC observations, reduction, and telluric correction method. In Section~\ref{2dcc} we describe the cross correlation and maximum likelihood analyses, and present the detection of the thermal spectrum of HD 88133 b. In Section~\ref{discussion} we discuss the implications of this result for the planet's atmosphere and for future observations.

\section{HIRES Observations and RV Analysis}
\label{RVs}
\begin{figure}[t]
\centering
\noindent\includegraphics[width=20pc]{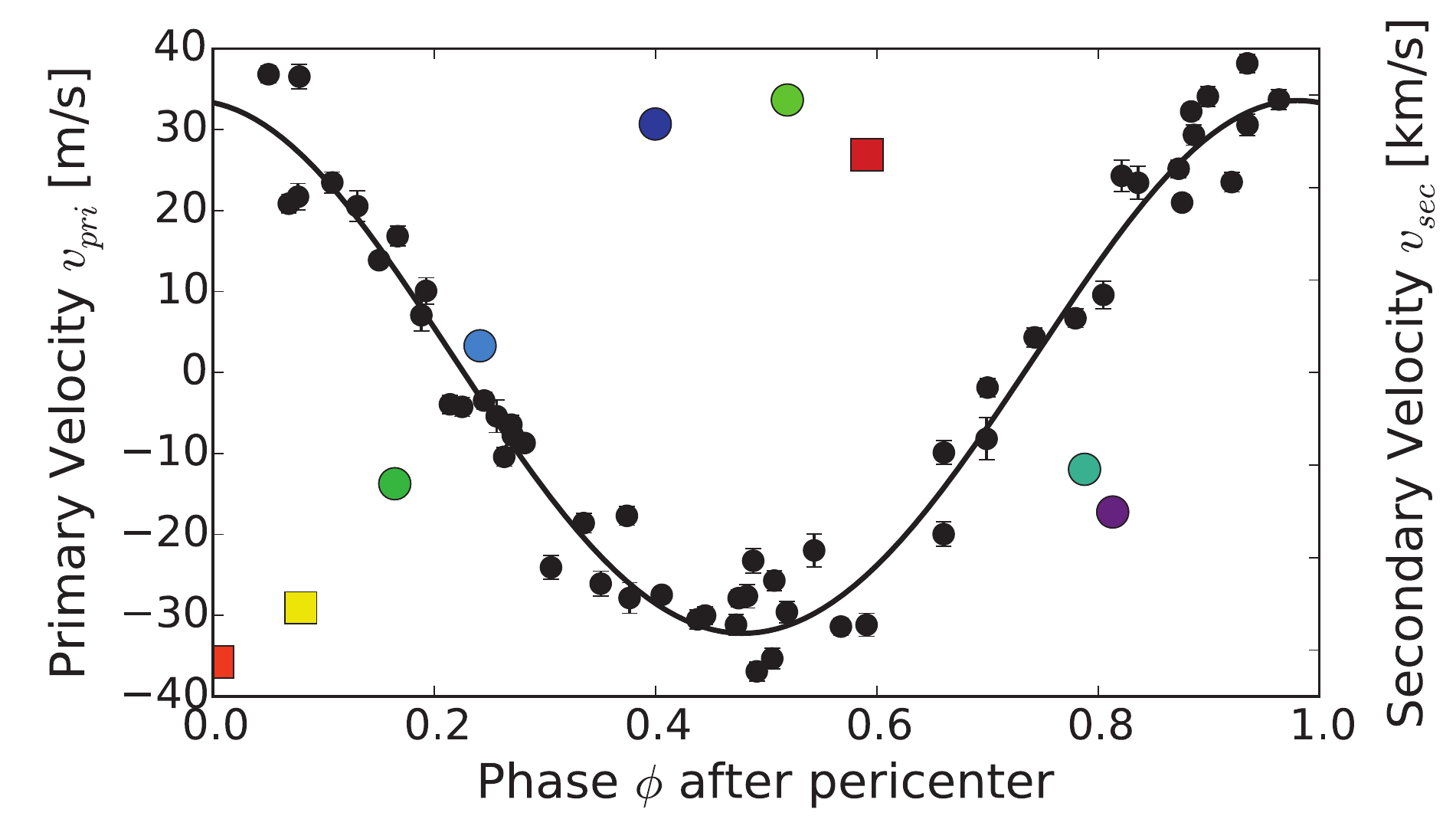}
\caption{RV data from the California Planet Survey with the best-fit stellar RV (primary velocity) curve overplotted in black. The colored points represent the NIRSPEC observations of this planet based on the observation phases and our qualitative expectations of their secondary velocities. The measure radial velocity of HD 88133 is 32.9$^{+1.03}_{-1.03}$ m/s}. In the course of this paper, we will show that the most likely value for the Keplerian orbital velocity of HD 88133 b is 40 $\pm$ 15 km/s.
\label{rvplot}
\end{figure}

The RV measurements of HD 88133 have been made under the purview of the California Planet Survey (CPS; \citealt{Howard2010}) with the HIgh Resolution Echelle Spectrometer (HIRES; \citealt{Vogt1994}) at the W.M. Keck Observatory. Seventeen RV measurements of HD 88133 were published in an earlier study \citep{Fischer2005}, and here we extend that data set to 55 individual RV measurements, having a baseline of eleven years (see Table~\ref{RVmsmts}). RV data are reduced with the standard CPS HIRES configuration and reduction pipeline \citep{Wright2004, Howard2010, Johnson2010}. Doppler shifts are recovered by comparison to an iodine absorption spectrum and a modeling procedure presented in \cite{Butler1996} and \cite{Howard2011}. Processed RV data are shown in Figure~\ref{rvplot}.

\begin{deluxetable}{lcc}[t]
\tablewidth{0pt}
\tablecaption{HD 88133 RV Measurements\tablenotemark{a}}
\tablehead{Julian Date  & Radial Velocity & $\sigma_{RV}$ \\
(- 2,440,000) & (m/s) & (m/s) } 
\startdata
13014.947812 &  -21.97 & 2.03  \\     
13015.947488 & 23.44 &   2.06 \\
13016.952546 &   20.55 &   1.91 \\
13044.088461 &   21.71 &   1.63 \\
13044.869410 & -24.07 &   1.46\\
13045.843414 &  -31.17 &   1.42\\
13046.081308 & -19.97 &    1.52 
\enddata
\tablenotetext{a}{The full set of RVs for this system is available as an electronic table online.} 
\label{RVmsmts}
\end{deluxetable}

The RV data are fit with a Markov Chain Monte Carlo technique following \cite{Bryan2016}. Eight free parameters (six orbital parameters: the velocity semi-amplitude $K$, the period of the orbit $P$, the eccentricity of the orbit $e$, the argument of periastron $\omega$, the true anomaly of the planet at a given time $f$, and the arbitrary RV zero point $\gamma$; a linear velocity trend $\dot\gamma$; and a stellar jitter term $\sigma_{jitter}$ as in \citealt{Isaacson2010}) having uniform priors contribute to the model $m$ and are simultaneously fit to the data $v$. We initiate the MCMC chains at the values published by \cite{Fischer2005}, allowing the chains to converge quickly and avoiding degeneracies in $e$ and $\omega$. The likelihood function is given by
\begin{equation}
\mathcal{L} = \frac{1}{\sqrt{2\pi}\sqrt{\sigma_{i}^{2} + \sigma_{jit}^{2}}}\exp\left(-0.5\left(\frac{(v-m)^2}{\sigma_{i}^{2} + \sigma_{jit}^{2}}\right)\right)
\end{equation}
where $\sigma_i$ is the instrument error and $\sigma_{jit}$ is the stellar jitter. The stellar jitter term is added in quadrature to the uncertainty value of each RV measurement. Best-fit orbital elements indicated by this analysis, as well as other relevant system parameters, are included in Table~\ref{systemproperties}, and the best-fit velocity curve is shown in the Figure~\ref{rvplot}. This represents a substantial improvement to the ephemeris originally published in \cite{Fischer2005}. We combine these values with the velocities derived by our NIRSPEC analysis described in Sections~\ref{methods} and~\ref{2dcc} to break the $M\sin(i)$ degeneracy for HD 88133 b.

\begin{deluxetable}{llc}[h]
\tablewidth{0pt}
\tabletypesize{\scriptsize}
\tablecaption{HD 88133 System Properties}
\tablehead{Property & Value & Reference} 
\startdata
\sidehead{\textbf{Stellar}}
Mass, $M_{st}$ & 1.18 $\pm$ 0.06 $M_{\sun}$ & (1)  \\
Radius, $R_{st}$ & 1.943 $\pm$ 0.064 $R_{\sun}$ & (2) \\
Effective temperaure, $T_{\mathrm{eff}}$ & 5438 $\pm$ 34 K & (1) \\
Metallicity, \feh & 0.330 $\pm$ 0.05 & (1) \\
Surface gravity, $\log g$ & 3.94  $\pm$ 0.11 & (1) \\
Rotational velocity, $v \sin i$ & 2.2 $\pm$ 0.5 & (1) \\
Systemic velocity, $v_{sys}$ & -3.45 $\pm$ 0.119 km/s & (3) \\
\textit{K} band magnitude, $K_{mag}$ & 6.2 & (4) \\
Velocity semi-amplitude, $K$ & 32.9$^{+1.03}_{-1.03}$ m/s & (5) \\
RV zero point, $\gamma$ & 3.08$^{+1.51}_{-1.47}$ m/s & (5) \\
Velocity trend, $\dot{\gamma}$ &  -0.0013$^{+0.0009}_{-0.0010}$ m/s/yr & (5) \\
Stellar jitter, $\sigma_{jitter}$ & 4.68$^{+0.51}_{-0.61}$ & (5) \\
\sidehead{\textbf{Planetary}}
Indicative mass, $M\sin(i)$ &  0.27$^{+0.01}_{-0.01}M_{Jup}$ & (5)\\
Mass, $M_p$ & 1.02$^{+0.61}_{-0.28}M_J$ & (6) \\
Inclination, $i$ & 15${^{+6}_{-5}}^{\circ}$ & (6) \\
Semi-major axis, $a$ & 0.04691 $\pm$ 0.0008 AU & (7)\\
Period, $P$ & 3.4148674$^{+4.57e-05}_{-4.73e-05}$ & (5) \\
Eccentricity, $e$ & 0.05$^{+0.03}_{-0.03}$ & (5) \\
Argument of periastron, $\omega$ &7.22${^{+31.39}_{-48.11}}^{\circ}$ & (5) \\
Time of periastron, $t_{peri}$ &  2454641.984$^{+0.293}_{-0.451}$ JD & (5) \\
Phase uncertainty, $\sigma_{f+\omega}$ & 6.34$^{\circ}$ & (5)\\
\enddata
\tablerefs{(1) \cite{Mortier2013}; (2) \cite{Torres2010}; (3) \cite{Chubak2012} (4) \cite{Wenger2000} (5) From HIRES measurements presented in Section~\ref{RVs}; (6) From NIRSPEC measurements presented in Sections~\ref{methods} and~\ref{2dcc}; (7) \cite{Butler2006}} 
\label{systemproperties}
\end{deluxetable}

\section{NIRSPEC Observations and Data Reduction}
\label{methods}
We pursue multi-epoch observations having planetary features shifted with respect to the star's spectrum and the Earth's atmosphere in order to develop techniques that can eventually be used on more slowly moving planets nearer the habitable zone. For nearly synchronously rotating hot Jupiters, near-infrared emission from the dayside is likely to be most readily detectable. This strategy is fundamentally different from that used at CRIRES to date since we do not allow the planet's signal to move across several pixels on the detector during one night of observations. 

\subsection{Observations}
\label{observations}
Data were taken on six nights (2012 April 1 and 3, 2013 March 10 and 29, 2014 May 14, 2015 April 8) in \textit{L} band and three nights (2015 November 21, 2015 December 1, and 2016 April 15) in \textit{K} band in an ABBA nodding pattern with the NIRSPEC instrument at the W.M. Keck Observatory \citep{McLean1998}. With a 0.4$''$x24$''$ slit NIRSPEC has an \textit{L} band resolution of 25,000 (30,000 in \textit{K} band). Individual echelle orders cover from 3.4038-3.4565/3.2567-3.3069/3.1216-3.1698/2.997-3.044 $\mu$m in the \textit{L} band and from 2.3447-2.3813/2.2743-2.3096/2.2085-2.2422/2.1464-2.1788/2.0875-2.1188/2.0319-2.0619 $\mu$m in the \textit{K} band.

\begin{figure}[t]
\centering
\noindent\includegraphics[width=20pc]{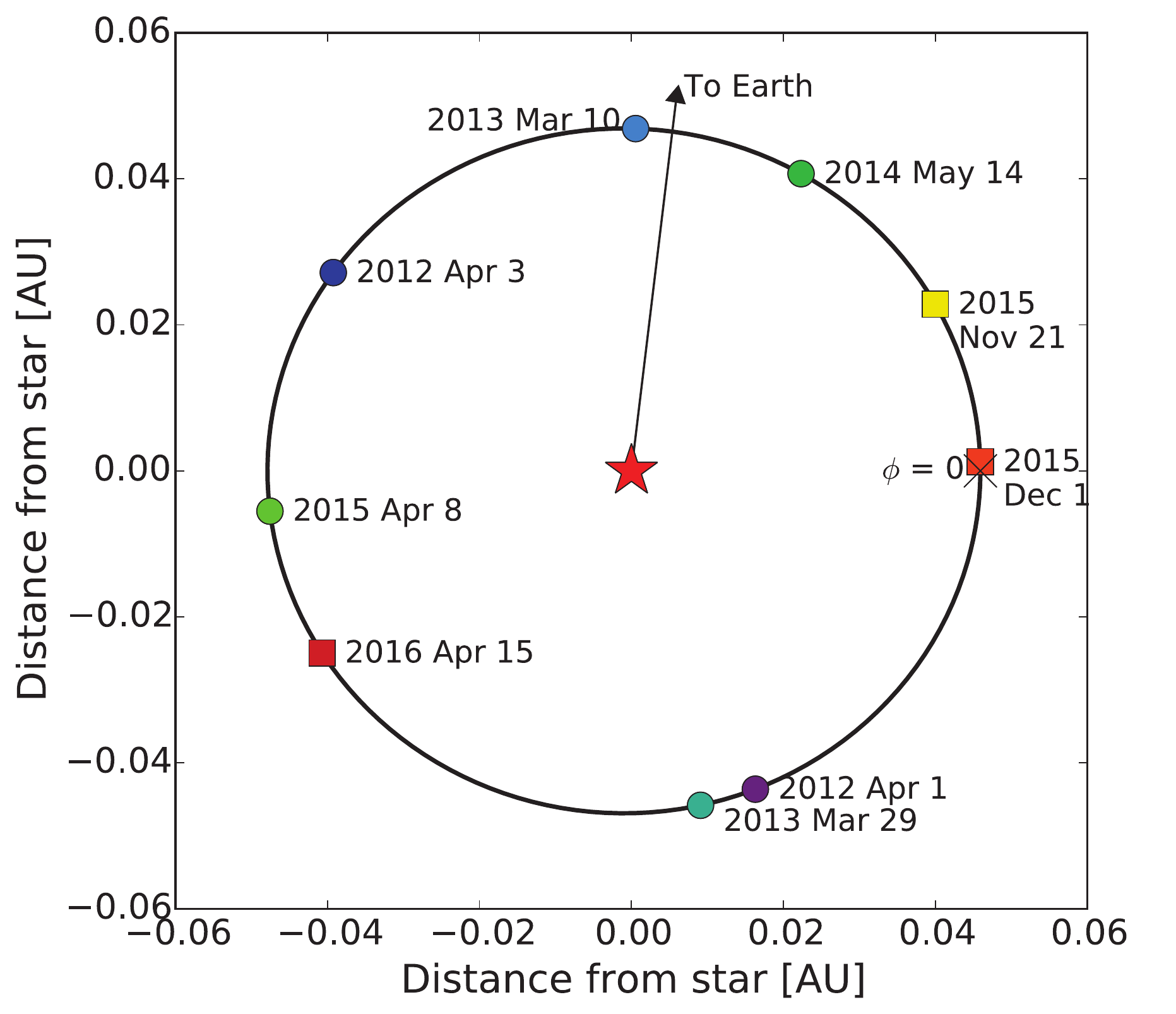}
\caption{Top-down schematic of the orbit of HD 88133 b around its star according to the orbital parameters derived by \cite{Fischer2005}, \cite{Butler2006}, and this work. Each point represents a single epoch's worth of NIRSPEC observations of the system. Circles indicate \textit{L} band observations and squares represent \textit{K} band observations. The black arrow represents the line of sight to Earth. }
\label{schematic}
\end{figure}

A schematic of the planet's orbit during our observations is given in Figure~\ref{schematic} assuming the best-fit orbital parameters from our HIRES RV analysis in Section~\ref{RVs}. Details of our observations are given in Table~\ref{observationtable}.

\begin{deluxetable*}{lcccccc}
\tablewidth{0pt}
\tablecaption{NIRSPEC Observations of HD 88133 b}
\tablehead{Date & Julian Date & Mean anomaly $M$ & True anomaly $f$ & Barycentric velocity $v_{bary}$ & Integration time & S/N$^{\tablenotemark{a}}_{\textit{L}, \textit{K}}$ \\
 & (- 2,440,000 days) &  (2$\pi$ rad) & (2$\pi$ rad) & (km/s) & (min) & }
\startdata
\sidehead{\textbf{\textit{L} band (3.0 - 3.4 $\mu$m)}}  
2012 April 1 & 16018.837 & 0.89 & 0.86 & -20.96 & 140 & 1680  \\
2012 April 3 & 16020.840 & 0.48 & 0.48 & -21.66 & 140 & 2219 \\
2013 March 10 & 16361.786 & 0.29 & 0.33 & -11.59 & 180 & 2472 \\
2013 March 29 & 16380.726 & 0.84 & 0.80 & -19.70 & 150& 1812 \\
2014 May 14 & 16791.796 & 0.18 & 0.22 & -29.27 & 120 & 1694 \\
2015 April 8 & 17120.835 & 0.50 & 0.50 & -23.01 & 160 & 2938 \\
\sidehead{\textbf{\textit{K} band (2.0 - 2.4 $\mu$m) } }
2015 November 21 & 17348.129 & 0.06 & 0.68 & 29.95 & 60 & 2701 \\
2015 December 1 & 17358.117 & 0.96 & 0.62 & 29.25 & 60 & 2823 \\
2016 April 15 & 17166.300 & 0.54 & 0.53 & -29.15 & 80 & 2466 \\
\enddata
\label{observationtable}
\tablenotetext{a}{S/N$_{\textit{L}}$ and S/N$_{\textit{K}}$ are calculated at 3.0 $\mu$m and 2.1515 $\mu$m, respectively. Each S/N calculation is for a single channel (i.e., resolution element) for the whole observation.} 
\end{deluxetable*}


\subsection{Extraction of 1-D Spectra}
\label{reduction}
\begin{figure}[t]
\centering
\noindent\includegraphics[width=20pc]{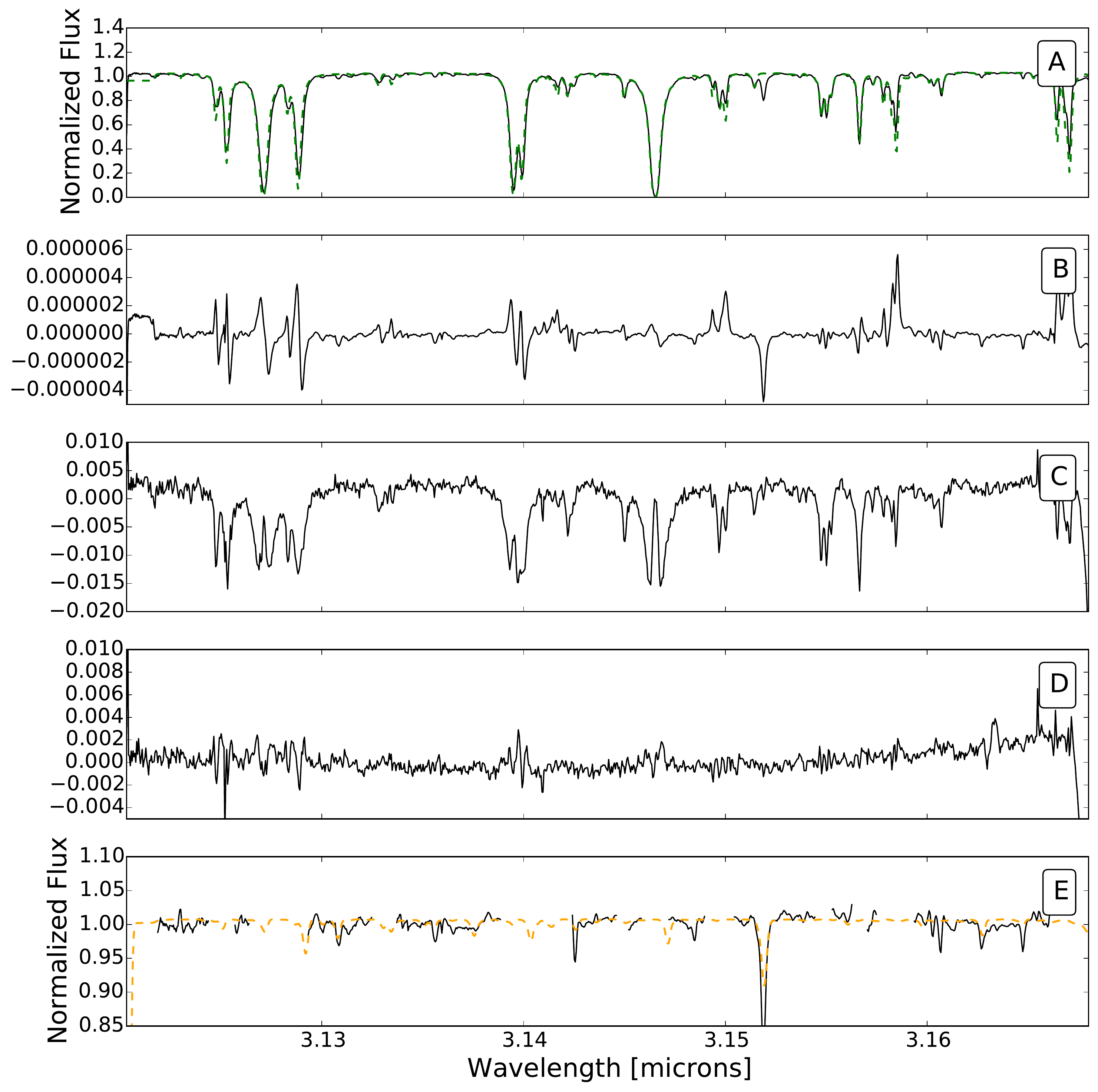}
\caption{The data reduction and telluric correction process. (A): One order of a reduced AB pair of HD 881333 data taken on 2013 March 29 in \textit{L} band, with a best-fit telluric spectrum overplotted with a green, dashed line.  (B): The first principal component in arbitrary units of this time series of data which encapsulates changes in the stellar spectrum as the air mass varies during the observation. (C): The second principal component in arbitrary units which describes changes in abundances of telluric species. (D) The third principal component  in arbitrary units which encompasses changes in plate scale. (E) Telluric-corrected data with the first five principal components removed shown in black. This is the data used for the cross-correlation analysis described in Section~\ref{2dcc}. Overplotted in orange is the stellar spectrum of HD 88133 adapted from the PHOENIX stellar library \citep{Husser2013}.}
\label{data}
\end{figure}

We flat field and dark subtract the data using a Python pipeline \`{a} la \cite{Boogert2002}. We extract one-dimensional spectra and remove bad pixels, and calculate a fourth-order polynomial continuum by fitting the data to a model telluric spectrum after the optimal source extraction. For \textit{L} band data, the wavelength solution (described to the fourth order as $\lambda = ax^3 + bx^3 + cx + d$, $x$ is pixel number and $a$, $b$, $c$, and $d$ are free parameters) is calculated by fitting the data to a model telluric spectrum. However, since telluric lines are generally weaker near 2 $\mu$m, the wavelength solution for \textit{K} band data is calculated by fitting the data to a combination of model telluric and stellar spectra (given that the stellar relative velocity is well known from optical data, and is later confirmed by the cross-correlation analysis described in Section~\ref{corr}). We use an adapted PHOENIX model for our model stellar spectrum in the \textit{K} band, as described in Section \ref{stellarmodel} \citep{Husser2013}. We show one order of reduced \textit{L} band spectra in the top panel of Figure~\ref{data}. We fit an instrument profile to the data and save it so that we may apply it to our stellar and planetary models. This instrument profile is similar to the formulation given in \cite{Valenti1995} and is parameterized as a central Gaussian with four left and four right satellite Gaussians, all with variable widths. Often, the best-fit widths of the third and fourth left and right satellite Gaussians are zero.

\subsection{Telluric Correction with Principal Component Analysis}
\label{pca}
In this work, we depart from our traditional methods of division by an atmospheric standard (typically an A star, c.f. \citealt{Boogert2002}) and/or line-by-line telluric correction (modeling atmospheric abundances with the TERRASPEC software; Bender et al., 2012) in favor of a more automated and repeatable technique: principal component analysis (PCA). This efficient method of telluric correction was also implemented by \cite{deKok2013} in their reduction of CRIRES data on HD 189733 b. PCA rewrites a data set in terms of its principal components such that the variance of the data set with respect to its mean or with respect to a model is reduced. The first principal component encapsulates the most variance; the second, the second most, etc. Over the course of an observation, the telluric components should vary the most as the air mass and atmospheric abundances change, and the planet lines should remain $\sim$constant. Note that we observe for only 2-3 hours at a time and at a lower resolution than CRIRES, so the planet lines do not smear. For a typical Keplerian orbital velocity of a hot Jupiter, we would have to observe for at least four hours to smear the planet lines across the NIRSPEC detector pixels. To remove the telluric lines and any other time-varying effects, we aim to isolate only the strongest principal components.

To perform the principal component analysis\footnote{\href{http://stats.stackexchange.com/questions/134282/relationship-between-svd-and-pca-how-to-use-svd-to-perform-pca}{PCA tutorial}, \ttfamily{http://stats.stackexchange.com/questions/\\134282/relationship-between-svd-and-pca-how-to-use-svd-to-\\perform-pca}}, we first reduce our data set in AB pairs and construct a data matrix $\mathbf{X}$ having $n$ rows and $m$ columns, where $n$ is the number of AB pairs and $m$ is the number of pixels (1024 for individual NIRSPEC echelle orders). We linearly interpolate sub-pixel shifts between nods when aligning the AB pairs on the matrix grid, and then calculate the residual matrix $\mathbf{R}$ according to 
\begin{equation}
\mathbf{R}_{ij} = \frac{\mathbf{X}_i - M}{\sigma_{j}}
\end{equation}
where $i$ is the row number, $j$ is the column number, $M$ is either the mean of $X_j$ or a telluric model, and $\sigma_{i}$ is the standard deviation of the values in column $j$. We guide our principal component analysis with a telluric model for $M$ (rather than the mean of  $X_j$) that uses baseline values for water vapor\footnote{\href{http://cso.caltech.edu/tau/}{Caltech Submillimeter Observatory}, \url{http://cso.caltech.edu/tau/}}, carbon dioxide\footnote{\href{ftp://aftp.cmdl.noaa.gov/products/trends/co2/co2_weekly_mlo.txt}{Earth System Research Laboratory at Mauna Loa}, \url{ftp://aftp.cmdl.noaa.gov/products/trends/co2/co2_weekly_mlo.txt}}, and methane\footnote{\href{ftp://ftp.cmdl.noaa.gov/data/greenhouse_gases/ch4/in-situ/surface/mlo/ch4_mlo_surface-insitu_1_ccgg_DailyData.txt}{Earth System Research Laboratory at Mauna Loa}, \url{ftp://ftp.cmdl.noaa.gov/data/greenhouse_gases/ch4/in-situ/surface/mlo/ch4_mlo_surface-insitu_1_ccgg_DailyData.txt}} abundances. For \textit{L} band data, baseline values for ozone from a reference tropical model are also included for the orders ranging from 3.12-3.17 and 3.26-3.31 $\mu$m. Next we calculate the covariance matrix $\mathbf{C}$ of our mean-normalized data such that 
\begin{equation}
\mathbf{C} = \frac{\mathbf{R^{T}R}}{n-1}.
\end{equation}
A singular value decomposition of the covariance matrix is then performed to find the principal components:
\begin{equation}
\mathbf{C} = \mathbf{USV^{T}}
\end{equation}
where $\mathbf{U}$ contains the left singular vectors (or the eigenvectors), $\mathbf{S}$ is a diagonal matrix of the singular values (or the eigenvalues), and $\mathbf{V}$ contains the right singular vectors. The first three eigenvectors, or principal components, for a 3.12-3.17 $\mu$m order taken on 2013 March 29  are shown in Figure ~\ref{data}. The first component recovers changes in the total systemic signal with air mass; the second encapsulates changes in abundances of telluric species, resulting in adjustments to line cores and wings; and the third describes changes to the plate scale. Higher order components typically reflect instrumental fringing and other small effects. 

We reconstruct the time-varying portion of each AB spectrum by combining the first $k$ principal components of the data set, given by $\mathbf{U}_{k}\mathbf{S}_{k}$, where $\mathbf{U}_{k}$ is the first $k$ columns of $\mathbf{U}$ and $\mathbf{S}_{k}$ is the $k$ x $k$ upper-left part of $\mathbf{S}$. Rank-$k$ data, $\mathbf{X}_{k}$, can be built as 
\begin{equation}
\mathbf{X}_k =  \mathbf{U}_{k}\mathbf{S}_{k}\mathbf{U}^{\mathbf{T}}_{k}.
\end{equation}
To produce a telluric-corrected spectrum, $\mathbf{X}_{corr,i}$,  each row $\mathbf{X}_i$ of $\mathbf{X}$ is divided by its corresponding un-mean normalized row in $\mathbf{X}_{k,i}$:
\begin{equation}
\mathbf{X}_{corr,i} = \frac {\mathbf{X}_i} {\mathbf{X}_{k,i}\sigma_{j}+M}
\end{equation}

Finally, we collapse the rows of data in $\mathbf{X}_{corr}$ and clip regions of substantial telluric absorption ($>$75\%). Depending on the order, anywhere between 30 and 60\% of the data is removed by clipping out strong features. This results in a single high signal-to-noise spectrum for each night of observations. The final telluric-corrected spectrum for 2013 March 29 is given in the final panel of Figure ~\ref{data}.

The version of PCA described here diverges from the approach outlined in \cite{deKok2013} which used PCA to determine the eigenvectors making up the light curves in each spectral channel. Our formulation calculates the eigenvectors comprising each observed spectrum. We also guide our principle component analysis with a telluric model. The equivalent for \cite{deKok2013} would have been to guide the PCA with vectors for air mass, water vapor content, etc.

For data taken in \textit{L} band, we find that PCA can reliably correct for the Earth's atmosphere for all orders of data. However, for \textit{K} band data, we cannot effectively remove the dense, optically-thick telluric forest of CO$_2$ lines in the order spanning from 2.03 - 2.06 $\mu$m, and we omit this wavelength range from subsequent analysis.

It is essential to determine the efficacy of PCA in removing telluric signatures and further ensure that we are not removing the planet's signal as well. Since $\sim$99.9\% of the variance is explained by the first principal component, we find that the following results are roughly consistent when different numbers of principal components are removed from the data. We calculate the percent variance removed by each component and, if we assume the planet signal is on the order of 10$^{-5}$ of the total signal, then for most cases we would have to remove about fifteen components to delete the planet signal. We have experimented with incrementally removing up to ten principal components as input to the subsequent analysis; the results and conclusions presented below use data with the first five principal components removed.

\section{NIRSPEC Data Analysis and Results}
\label{2dcc}
We run a two-dimensional cross-correlation analysis (TODCOR algorithm; \citealt{Zucker1994}) to find the optimum shifts for the stellar and planetary spectra entwined in our telluric- and instrument-corrected data. This requires accurate model stellar and planet spectra. 

\subsection{Model Stellar Spectrum}
\label{stellarmodel}
Our synthetic stellar spectrum is a PHOENIX model \citep{Husser2013} interpolated to match the published effective temperature $T_{\mathrm{eff}}$, surface gravity  $\log g$, and metallicity \feh ~of HD 88133 listed in Table \ref{systemproperties}. As HD 88133 has a $v \sin i$ $<$ 5 km/s, instrumental broadening dominates, and we convolve the stellar model with the instrumental profile calculated in Section \ref{reduction}.

\subsection{Model Planetary Spectrum}
\begin{figure}[t]
\centering
\noindent\includegraphics[width=20pc]{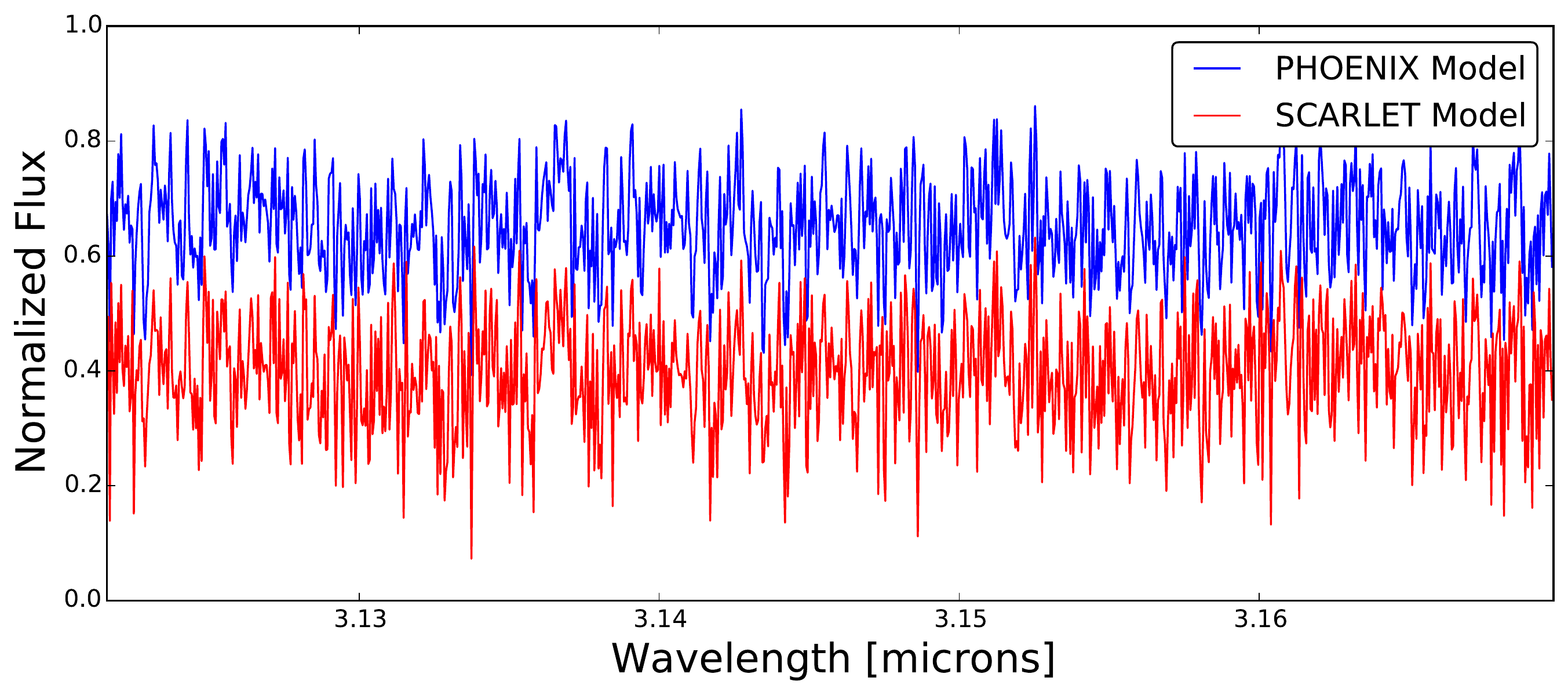}
\caption{Forward models for the planetary atmosphere of HD 88133 b produced by the PHOENIX and SCARLET models drawn at instrument resolution. Note that the flux calculated by the SCARLET model is shifted downward by 0.3 for clarity. Features shown here are principally due to water vapor. The correlation coefficient between these two models at zero-lag is 0.92.}
\label{planet}
\end{figure}

We have computed the high-resolution thermal emission spectrum of HD 88133 b using both the SCARLET \citep{Benneke2015} and PHOENIX \citep{Barman2001, Barman2005} frameworks. An example of one order of our \textit{L} band planet models is shown in Figure ~\ref{planet}. Both models compute the thermal structure and equilibrium chemistry of HD 88133 b given the irradiation provided by the host star. Models are computed for a cloud-free atmosphere with solar elemental composition \citep{Asplund2009} at a resolving power of R $>$ 250K, and assume perfect heat redistribution between the day and night sides. The model spectra are subsequently convolved with the instrumental profile derived in Section~\ref{reduction} (Figure~\ref{planet}). We find consistent results for both models despite minor differences in the molecular line lists used.

The SCARLET model considers the molecular opacities of $\mathrm{H_{2}O}$, $\mathrm{CH_{4}}$, $\mathrm{NH_{3}}$, HCN, $\mathrm{CO}$, and $\mathrm{CO_{2}}$ and $\mathrm{TiO}$ from the high-temperature ExoMol database \citep{Tennyson2012}, and $\mathrm{O_{2}}$, $\mathrm{O_{3}}$, $\mathrm{OH}$, $\mathrm{C_{2}H_{2}}$, $\mathrm{C_{2}H_{4}}$, $\mathrm{C_{2}H_{6}}$, $\mathrm{H_{2}O_{2}}$, and $\mathrm{HO_{2}}$ from the HITRAN database \citep{Rothman2009}. Absorption by the alkali metals (Li, Na, K, Rb, and Cs) is modeled based on the line strengths provided in the VALD database \citep{Piskunov1995} and $\mathrm{H_{2}}$-broadening prescription provided in \cite{Burrows2003}. Collision-induced broadening from $\mathrm{H_{2}}/\mathrm{H_{2}}$ and $\mathrm{H_{2}/He}$ collisions is computed following \cite{Borysow2002}. 

Unlike SCARLET, PHOENIX is a forward modeling code that converges to a solution based on traditional
model atmosphere constraints (hydrostatic, chemical, radiative-convective, and local thermodynamic equilibrium) for an assumed elemental composition.  The PHOENIX model uses similar opacities as SCARLET (for example, most of the latest linelists from ExoMol and HITRAN).  Additional line data for metal-hydrides come from \cite{Dulick2003} (and references therein). 
 Broadening of alkali lines follows \cite{Allard2003}.  These differences are of only minor importance for this study because SCARLET and PHOENIX use the same water linelist and water opacity dominates the spectral features across the spectral range of our observations (Figure~\ref{planet}). 

Based on the effective temperatures of the planet and the star, the photometric contrast $\alpha_{phot}$ (defined as the ratio of the planet flux to the stellar flux) is on the order of 10$^{-4}$. This is also a rough upper bound for the spectroscopic contrast $\alpha_{spec}$. Since the cross-correlation analyses described in Section~\ref{corr} are not very sensitive to contrast ratios, varying the value of $\alpha_{spec}$ does not change our conclusions on the radial velocity of the planet, and thus the system inclination \citep{Lockwood2014}. However, the specific value of $\alpha_s$ does affect our conclusions on the composition and structure of the planet's atmosphere. That is, the overall velocity structure of the cross-correlation surface is not much affected by $\alpha_{spec}$, though the size and structure of the final maximum likelihood peak near the planet's signature will be.

\subsection{Two-Dimensional Cross Correlation}
\label{corr}
Each order of data for each epoch is cross-correlated with the model predictions to determine the cross-correlation function (CCF) using the TODCOR algorithm \citep{Zucker1994}. This calculation results in a two-dimensional array of correlation values for shifts in the velocities of the star and planet. 

In testing our models, we find that the correlation coefficient between our stellar and planet models for the two orders spanning from 2.31 - 2.38 $\mu$m is generally an order of magnitude higher than the same correlation coefficient in any other order in the \textit{L} or \textit{K} band, and so we omit them from this study. This behavior is due to the strong correlation between stellar and planetary CO at R=30,000. HD 88133 has an effective temperature of 5438 K and its CO band (and especially the CO band head at 2.295 $\mu$m) is extremely prominent. The \textit{K} band data analysis that follows is only performed on the three orders spanning from 2.10 - 2.20 $\mu$m. The main absorbers in this region include carbon dioxide and water vapor.

\begin{figure}[t]
\centering
\noindent\includegraphics[width=20pc]{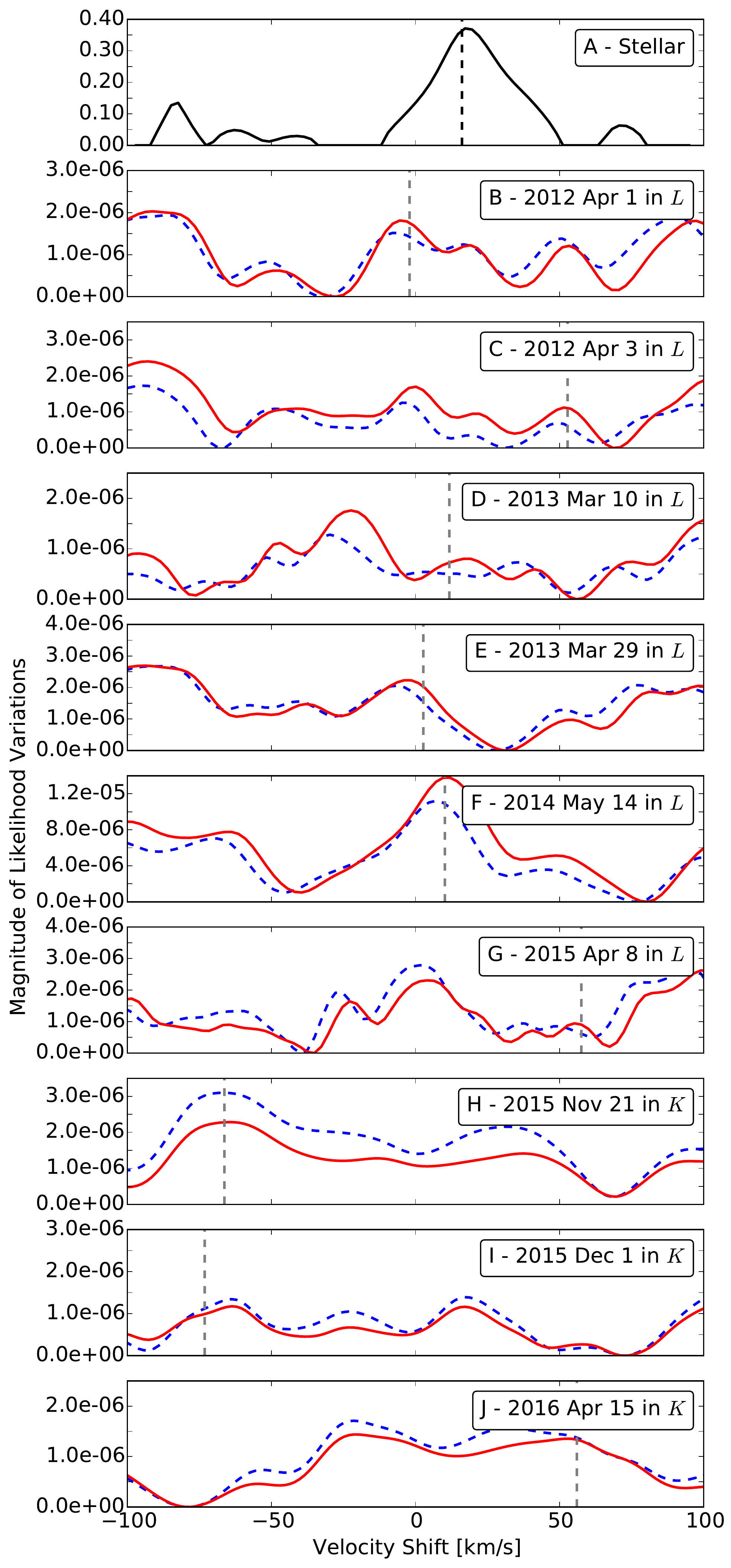}
\caption{Maximum likelihood functions for each observational epoch. (A): Maximum likelihood function of the stellar velocity shift of data taken on 2013 March 29. The black vertical dashed line indicates the expected stellar velocity shift. (B) - (G): Maximum likelihood function of $v_{sec}$ for \textit{L} band data from 3.0-3.4 $\mu$m taken on 2012 April 1 and 3, 2013 March 10 and 29, 2014 May 14, and 2015 April 8, respectively. (H) - (J): Maximum likelihood function of $v_{sec}$ for \textit{K} band data from 2.10-2.20 $\mu$m taken on 2012 November 21, 2015 December 1, and 2016 April 15, respectively. Note in (B) - (J), the blue dashed curve shows the maximum likelihood function for the PHOENIX model, the red curve shows the maximum likelihood function for the SCARLET model, and the grey vertical dashed lines indicate the planetary velocity shift on that date given an orbital solution having $K_P$ = 40 km/s. Based on $\sigma_{f+\omega}$, the error on the calculated planetary velocity shift is about 1.2 km/s.}
\label{sxcorr}
\end{figure}

For each night's observation, we combine the CCF's of each order with equal weighting to produce the maximum likelihood curves shown in Figure~\ref{sxcorr}.  \cite{Lockwood2014} showed that the cross-correlation function is proportional to the log of the likelihood. At each epoch, we can easily retrieve and confirm the stellar velocity, as shown by the single strong peak in panel A of Figure ~\ref{sxcorr}. The stellar velocity is dominated by the barycentric velocity at the time of observation and the systemic velocity of HD 88133. We are insensitive to the reflex motion of the star, which is on the order of 0.01 km/s. 

The retrieval of the planet velocity $v_{sec}$ is more complex, as evidenced by the multiple peaks in panels B-J  of Figure \ref{sxcorr}, and requires combining the data from multiple epochs. Though there are many peaks and troughs in the maximum likelihood curves produced for each night's observation (Figure~\ref{sxcorr}), only one peak per night represents the properly registered correlation of the data with the model planetary spectrum. This is not to say that most of the maximum likelihood peaks in Figure~\ref{sxcorr} are spurious; rather, they are the results of unintended systematic structure in the cross-correlation space.

\subsection{Planet Mass and Orbital Solution}
\begin{figure}[t]
\centering
\noindent\includegraphics[width=20pc]{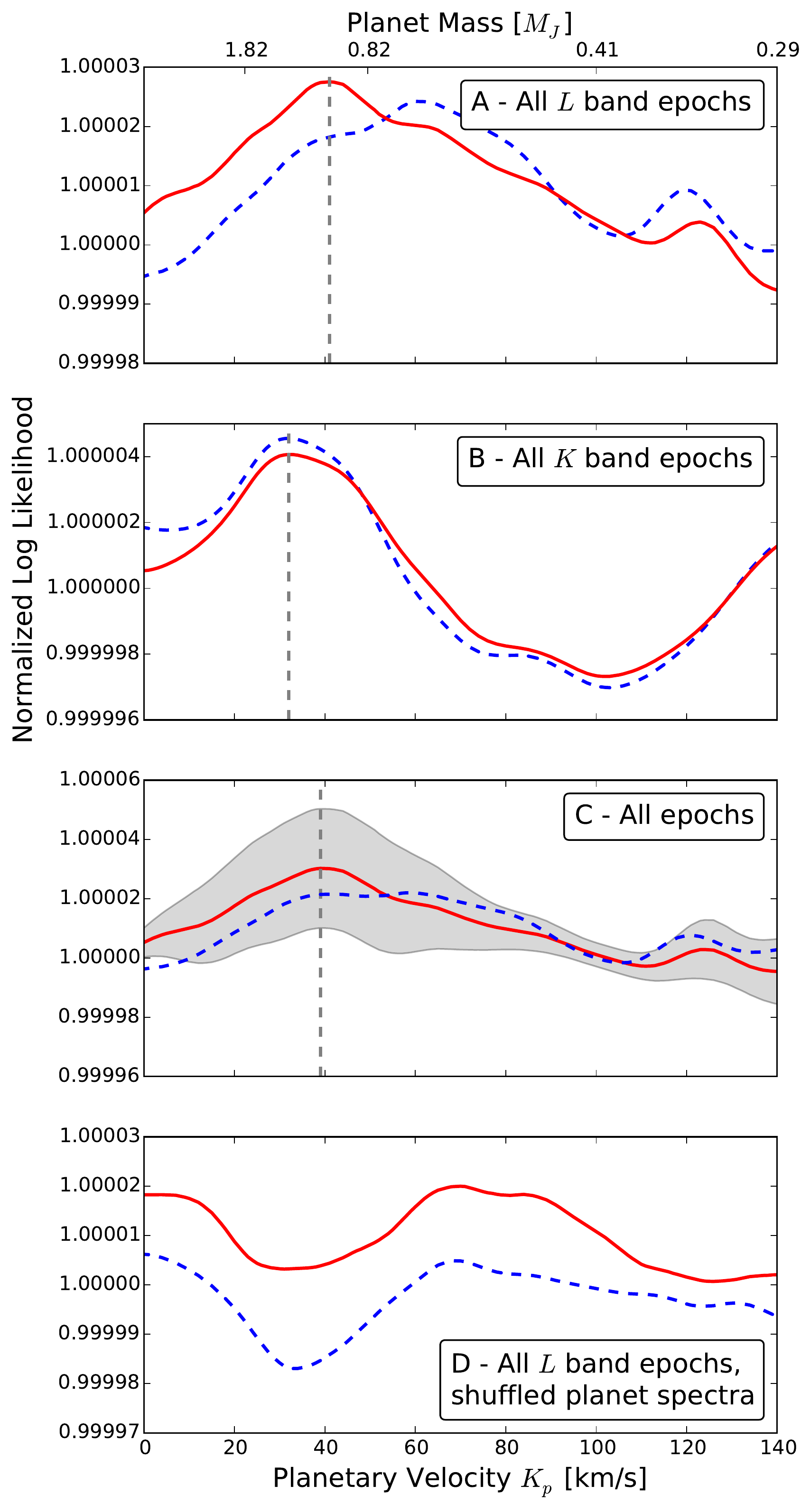}
\caption{Normalized log likelihood as a function of Keplerian orbital velocity $K_P$. Note that the vertical axes cannot be directly compared, and the color scheme is the same as Figure~\ref{sxcorr}. (A): Normalized log likelihood curve for six nights of \textit{L} band data from 3.0-3.4 $\mu$m. (B): Normalized log likelihood curve for three nights of \textit{K} band data from 2.10-2.20 $\mu$m. (C): Normalized log likelihood for all epochs and orders of NIRSPEC data used in panels (A) and (B). The grey region represents the one-sigma error bars determined by jackknife sampling for data cross-correlated with the SCARLET model. (D): Normalized log likelihood curve for six nights of \textit{L} band data cross-correlated with shuffled planetary spectra.}
\label{kp}
\end{figure}

The orbit of HD 88133 b is slightly eccentric, and we calculate the velocity $v_{sec}$ of the planet for a given epoch as a function of its true anomaly $f$ according to
\begin{equation}
\label{vf}
v_{sec}(f) = K_{p}(\cos(f+\omega)+e\cos\omega) + \gamma
\end{equation}
where $K_p$ is the planet's Keplerian orbital velocity, $\omega$ is the longitude of periastron measured from the ascending node, $e$ is the eccentricity of the orbit, and $\gamma$ is the combined systemic and barycentric velocities at the time of the observation. We test a range of orbital velocities from -150 to 150 km/s in steps of 1 km/s, and in turn test a range of planet masses and orbital inclinations. Figure~\ref{kp} shows the maximum log likelihood versus the planet's Keplerian orbital velocity.

When the six maximum likelihood curves produced from \textit{L} band data (panels B-G in Figure~\ref{sxcorr}) are combined with equal weighting according to Eq.~\ref{vf}, we see that the most likely value for the radial projection of the planet's Keplerian orbital velocity $K_P$ is 41 $\pm$ 16 km/s. These error bars are calculated as in \cite{Lockwood2014} by fitting a Gaussian to the likelihood peak and reporting the value of $\sigma$, assuming all the points on the maximum likelihood curve are equally weighted. We deduce that the peak in the likelihood curve based on the PHOENIX model at $\sim$60 km/s does not represent the planet's velocity, and we prove this in Section~\ref{tests}. The same calculation applied to the three nights of \textit{K} band data (panels H-J in Fig.~\ref{sxcorr}) suggests that $K_P$ is 32 $\pm$ 12 km/s. 

The combination of all nine nights of data yields $K_P$ = 40 $\pm$ 15 km/s. It is this value of $K_P$ that we use to calculate the secondary velocity curve shown in the bottom panel of Figure~\ref{schematic}. The peak near $K_P$ = 40 km/s is consistent between the two planet models cross-correlated with the data.  Here, given the full suite of data, we calculate error bars on the individual points with jackknife sampling. One night's worth of data is removed from the sample and the maximum likelihood calculation is repeated. The standard deviation of each point amongst the resulting eight maximum likelihood curves is proportional to the error bars on each point on the maximum likelihood curve.

We note that error bars calculated by jackknife sampling and shown in Figure~\ref{kp} are merely an estimate. In fact, for the Gaussian fit, the reduced chi-squared value (chi-squared divided by the number of degrees of freedom) is 0.1, indicating that the error bars are overestimated. This can be explained by the fact that there is high variance between jackknife samples, driving a high standard deviation and therefor large error bars. To examine this behavior further, we fit a Gaussian distribution (indicating the presence of a planetary signal) and a flat line (indicating no planetary signal) and determine the significance of the signal. As in \cite{Kass1995}, we define the Bayes factor $B$ to be the ratio of likelihoods between two models, in this case the likelihood of the Gaussian distribution compared to the likelihood of the straight line. $2\mathrm{ln}B$ must be greater than 10 for a model to be very strongly preferred.

For the Gaussian distribution compared the the straight line, $2\mathrm{ln}B$ is nearly 430, indicating the the Gaussian approximation to the signal at 40 km/s is significantly stronger than a flat line. Even if our error bars are overestimated, they are likely not overestimated by a factor of 100. Combining $K_P$ with the parameters given in Table~\ref{systemproperties}, a $K_P$ of 40 $\pm$ 15 km/s implies that the true mass of HD 88133 b is 1.02$^{+0.61}_{-0.28}M_J$ and its orbital inclination is 15${^{+6}_{-5}}^{\circ}$.

Note that the values of $v_{sec}$ implied by the most likely value of $K_P$ often, but do not always, correspond with peaks in the maximum likelihood curves for each night, as indicated by the vertical grey dashed lines in Figure~\ref{sxcorr}. Especially for nights having a small line-of-sight velocity, planetary lines may be lost in the telluric and/or stellar cross correlation residuals. 

\section{Discussion}
\label{discussion}

\subsection{Tests of the Orbital Solution}
\label{tests}
We first check our detection of HD 88133 b's emission spectrum at a $K_P$ radial velocity projection of 36 km/s by varying the spectroscopic contrast $\alpha_{spec}$ uniformly with orbital phase. We tested nine values of $\alpha_{spec}$ from 10$^{-7}$ to 10$^{-3}$, and find the maximum likelihood peak near 40 km/s is robust for $\alpha_{spec}$ $\ge$ 10$^{-5.5}$.

We create a ``shuffled" planetary model by randomly rearranging chunks of each planetary atmosphere model. If the maximum likelihood peak at 36 km/s 
is real, then cross correlating our data with a ``shuffled" planetary model (which has no coherent planet information) should show little to no peak at the expected $K_P$. And, indeed, the data-``shuffled" planetary model cross-correlation shows no peak at 40 km/s while the peak at $\sim$ 60 km/s remains for the PHOENIX model (see Panel D of Figure~\ref{kp}).

We also check our results by varying the orbital elements of the system. We obtain roughly the same values for Keplerian orbital velocity (within the error bars) for various combinations of eccentricities down to $\sim$ 0 and arguments of perihelion within 20$^{\circ}$ of the reported value. Our results are most sensitive to these orbital elements as they affect the calculations of the true anomaly and secondary velocity, and therefore the positions of the dashed vertical lines in each epoch's maximum likelihood curve shown in Figure \ref{sxcorr}. Even with a different ephemeris, so long as the dotted vertical lines are near their current peaks, we obtain a comparable final result for the Keplerian orbital velocity of the system.

We note that as HD 88133 b's orbit is slightly eccentric, it is not truly tidally locked. Calculations following \cite{Hut1981} suggest that the planet spins about 10\% faster than synchronous. Our strategy prefers that the planet be tidally locked so that as much of the planet's (dayside) emission as possible is captured when the planet has a high line-of-sight velocity relative to the star. 

Finally, we have reprocessed the tau Boo b data published by \cite{Lockwood2014} with the methods presented in Section~\ref{methods}. The specific departure from \cite{Lockwood2014} is the use of principal component analysis to correct for telluric features in the data (Section~\ref{pca}). Our analysis of five nights of \textit{L} band data recovers a projected Keplerian orbital velocity of 121 $\pm$ 8 km/s, a mass of 5.39$^{+0.38}_{-0.24}M_J$, and an orbital inclination of 50${^{+3}_{-4}}^{\circ}$ for tau Boo b. This is in good agreement with \cite{Lockwood2014} ($K_P$=111$\pm$5 km/s) as well as \cite{Brogi2012} ($K_P$=110$\pm$3.2 km/s) and \cite{Rodler2012} ($K_P$=115$\pm$11 km/s), thereby validating our PCA-based methods.

\subsection{Observation Notes}
The peak in log likelihood produced solely from \textit{L} band data (panel A of Figure~\ref{kp}) is an order of magnitude larger than the peak produced solely from \textit{K} band data, though the values of $K_P$ preferred by each data set are consistent. Though the single channel signal-to-noise ratio for our data is greater in the \textit{K} band than in the \textit{L} band, we only observe HD 88133 in \textit{K} for three nights and only use three orders of data in our final cross-correlation analysis whereas we observe in \textit{L} for six nights and use all four orders of data. Furthermore, six nights of \textit{L} band observations on Keck yields an aggregate shot noise of $\sim$ 800,000 for all epochs and all wavelength bins, suggesting that the detection of a planet having a spectroscopic contrast down to 10$^{-5}$ is feasible. 

Additionally, we attempt to detect the planet's CO band near 2.295 $\mu$m, and find that effectively separating the stellar spectrum from the planet's is a complex process. Future observations should consider avoiding the CO band head and focusing on the CO comb (low- to moderate- angular momentum P and R branches) itself, particularly the regions between the stellar lines where shifted planetary CO should be present. These intermediate regions have $\Delta$v $\sim$ 60 km/s, which is certainly sufficient for the detection of shifted planetary CO and especially so given the high resolution of the next generation of cross dispersed infrared echelle spectrographs, including iSHELL and the upgraded CRIRES and NIRSPEC instruments.

\subsection{The Spectrum of HD 88133}
We took the opportunity during our reprocessing of the tau Boo b data to evaluate the required accuracy of the stellar model. The original \cite{Lockwood2014} analysis was performed with a stellar spectrum using a MARCS solar model with adjustments made to specific line parameters. We re-run the analysis with a PHOENIX model for tau Boo. The correct stellar velocity at each observation epoch is still recovered and the final result for the planet's $K_P$ remains unchanged. This suggests that, for the sake of detecting the planet's spectrum and thus the planet's velocity, a detailed model of the star is not necessarily required; however a refined stellar spectrum will be critical for learning about the planet's atmosphere in detail.

\subsection{The Atmosphere of HD 88133 b}
Acquisition of both \textit{L} and \textit{K} band NIRSPEC data provides a unique opportunity to constrain the atmosphere of HD 88133 b. Generally, the spectroscopic contrast is dominated by water vapor in the \textit{L} band and by carbon monoxide and other species in the \textit{K} band. Ultimately, the shape of the log likelihood peak in Figure~\ref{kp} will provide information about the structure of the atmosphere (e.g., \citealt{Snellen2010} and \citealt{Brogi2016}) and even information about the planet's rotation rate (e.g., \citealt{Snellen2014} and \citealt{Brogi2016}). 
We do not presently consider orbital phase variations in atmosphere dynamics, radiative transfer, and the resulting spectral signatures from the day to night sides of the hot Jupiter. We plan to examine detailed properties of HD 88133 b's atmosphere in a future study. 

\section{Conclusion}
We report the detection of the emission spectrum of the non-transiting exoplanet HD 88133 b using high resolution near-infrared spectroscopy. This detection is based on the combined effect of thousands of narrow absorption lines, predominantly water vapor, in the planet's spectrum. We find that HD 88133 b has a Keplerian orbital velocity of  40 $\pm$ 15 km/s, a true mass of 1.02$^{+0.61}_{-0.28}M_J$, and a nearly face-on orbital inclination of 15${^{+6}_{-5}}^{\circ}$. 

Direct detection of hot Jupiter atmospheres via this approach is limited in that it cannot measure the {\it absolute} strengths of molecular lines, relative to the photometric contrast. Thus, this method will yield degeneracies between the vertical atmospheric temperature gradients and absolution molecular abundance ratios, but the relative abundances of species should be better constrained. For transiting planets having Spitzer data, it should be possible to better measure absolute abundances by comparing Spitzer eclipse depths and the output of our cross-correlation analyses using various planetary atmosphere models.

With the further refinement of this technique and with the improved future implementation of next-generation spectrometers and coronagraphs, especially on the largest optical/infrared telescopes, we are optimistic that this method may be extended to the characterization of terrestrial atmospheres at Earth-like semi-major axes. This paper shows progress in that direction by presenting an algorithm capable of removing telluric lines whilst preserving planet lines even if the planet does not move significantly during the observations.

\acknowledgments{The authors would like to thank Heather Knutson for helpful discussions throughout the preparation of this manuscript. The authors wish to recognize and acknowledge the very significant cultural role and reverence that the summit of Mauna Kea has always had within the indigenous Hawaiian community.  We are most fortunate to have the opportunity to conduct observations from this mountain. The data presented herein were obtained at the W.M. Keck Observatory, which is operated as a scientific partnership among the California Institute of Technology, the University of California and the National Aeronautics and Space Administration. The Observatory was made possible by the generous financial support of the W.M. Keck Foundation. This work was partially supported by funding from the NSF Astronomy \& Astrophysics and NASA Exoplanets Research Programs (grants AST-1109857 and NNX16AI14G, G.A. Blake P.I.), and the Center for Exoplanets and Habitable Worlds, which is supported by the Pennsylvania State Unviersity, the Eberly College of Science, and the Pennsylvania Space Grant Consortium. Basic research in infrared astrophysics at the Naval Research Laboratory is supported by 6.1 base funding. Finally, we thank an anonymous reviewer for helpful insights which improved the content of this paper.}

\end{document}